\documentclass[12pt]{article}
\usepackage{ifthen}
\newboolean{PRL}
\setboolean{PRL}{false}  
\newcommand{\PRL}[2]{\ifthenelse{\boolean{PRL}}{#1}{#2}}
\newboolean{Brit}
\setboolean{Brit}{true}  
\newcommand{\spell}[2]{\ifthenelse{\boolean{Brit}}{#1}{#2}}
\PRL{}{
\setlength{\oddsidemargin}{0in}
\setlength{\evensidemargin}{0in}
\setlength{\topmargin}{-4pc}
\setlength{\headheight}{1pc}
\setlength{\headsep}{2pc}
\setlength{\textwidth}{6.5in}
\setlength{\textheight}{9in}
\addtolength{\parskip}{1mm}
\clubpenalty 10000
\widowpenalty 10000
\sloppy
}
\newcommand{\jbox}{\hfill\rule{2mm}{2mm}}

\def\ket#1{\mbox{$| #1 \rangle$}}
\def\Tr{\mbox{Tr}}

\def\qed{\PRL{$\Box$}{\rule{0.5em}{0.809em}}}
\def\01{\{0,1\}}
\def\half{\textstyle{1 \over 2}}
\def\pihalf{\textstyle{\pi \over 2}}
\newcommand{\squash}[1]{\raisebox{0.04ex}[0pt][0pt]{\small$\textstyle #1$}}
\def\shalf{\squash{\frac{1}{\sqrt{2}}}}

\def\xor{\oplus}

\def\ni{\noindent}
\def\nl{\newline}
\def\ii{\hspace*{8mm}}
\def\xx{x^{\prime}}
\def\yy{y^{\prime}}
\def\aa{a^{\prime}}
\def\bb{b^{\prime}}
\def\bell{\ket{\Phi^{+}}_{\! AB}}
\def\epr{\ket{\Psi^{-}}_{\! AB}}
\def\belln{\ket{\Phi^{+}}^{\otimes n}_{\! AB}}
\def\entang{\ket{\Psi}_{\! AB}}

\begin{document}

\PRL{}{\date{15 January 1999}}

\title{The cost of exactly simulating quantum entanglement with classical 
communication}

\PRL{
\author{Gilles Brassard$^1$\thanks{Email: {\tt brassard@iro.umontreal.ca}},
Richard Cleve$^2$\thanks{Email: {\tt cleve@cpsc.ucalgary.ca}},
Alain Tapp$^1$\thanks{Email: {\tt tappa@iro.umontreal.ca}}}
\address{$^1$ D\'epartement IRO, Universit\'e de Montr\'eal, C.P.
6128,  Succursale Centre-Ville, Montr\'eal, Qu\'ebec, Canada H3C 3J7\\
$^2$ Department of Computer Science, University of Calgary,
Calgary, Alberta, Canada T2N 1N4}}
{
\author{
Gilles Brassard\,%
\thanks{\,Supported in part by Canada's {\sc nserc}, Qu\'ebec's {\sc fcar}
and the Canada Council.}\\
{\protect\small\sl Universit\'e de Montr\'eal\/}\,%
\thanks{\,D\'epartement IRO, Universit\'e de Montr\'eal,
C.P. 6128, succursale centre-ville, Montr\'eal (Qu\'ebec), 
Canada H3C 3J7.  Email: \{{\tt brassard},{\tt tappa}\}%
@{\tt iro.umontreal.ca}.}
\and
Richard Cleve\,%
\thanks{\,Supported in part by Canada's {\sc nserc}.}\\
{\protect\small\sl University of Calgary\/}\,%
\thanks{\,Department of Computer Science, University of Calgary,
Calgary, Alberta, Canada T2N 1N4.
Email: {\tt cleve}@{\tt cpsc.ucalgary.ca}.}
\and
Alain Tapp\,\footnotemark[3]
\\
{\protect\small\sl Universit\'e de Montr\'eal\/}\,\footnotemark[2]
}
}

\maketitle
\PRL{}{\thispagestyle{empty}}

\begin{abstract}
We investigate the amount of communication that must augment 
classical local hidden variable models in order to simulate the 
\spell{behaviour}{behavior} of entangled quantum systems.
We consider the scenario where a bipartite measurement is given from a set 
of possibilities and the goal is to obtain exactly the same correlations 
that arise when the actual quantum system is measured.
We show that, in the case of a single pair of qubits in a Bell state,
a constant number of bits of communication is always sufficient---regardless 
of the number of measurements under consideration.
We also show that, in the case of a system of $n$ Bell states, 
a constant times $2^n$ bits of communication are necessary.
\end{abstract}

\PRL{}{}

\PRL{}{\section{Introduction}}

Bell's celebrated theorem \cite{Bell} shows that certain scenarios involving 
bipartite quantum measurements result in correlations that are impossible 
to simulate with a classical system if the measurement events are 
space-like separated.
If the measurement events are time-like separated then classical 
simulation is possible, at the expense of some communication.
Our goal is to quantify the required amount of communication.

\PRL{This}{The issue that we are addressing} is part of the broader question 
of how quantum information affects various resources required to perform 
tasks in information processing.
A~two-way classical communication channel between two separated parties 
can be regarded as a {\em resource}, and a natural goal is for two 
parties to produce classical information satisfying a specific 
stochastic property.
One question is, if the parties have an {\em a priori\/} supply of quantum 
entanglement, can they accomplish such goals with less classical 
communication than necessary in the case where their {\em a priori\/} 
information consists of only classical probabilistic information?
And, if so, by how much?
Our question is, to what extent does the fundamental 
\spell{behaviour}{behavior} of an entangled quantum system itself 
provide savings, in terms of communication compared with classical 
systems?

Imagine a scenario involving two ``particles'' that may have been 
``together'' (and interacted) at some previous point in time, but are 
``separated'' (in a sense which implies that they can no longer 
interact) at the present time.
Suppose that a measurement is then arbitrarily selected and performed 
on each particle (not necessarily the same measurement on both particles).
If the underlying physics governing the \spell{behaviour}{behavior} of the
system is  ``classical'' then the \spell{behaviour}{behavior} of such a 
system could be based on  correlated random variables (usually called 
``local hidden variables''),  reflecting the possible results of a previous 
interaction.
If no communication can occur between the components at the time when 
the measurements take place then this imposes restrictions on the 
possible \spell{behaviour}{behavior} of such a system.
In fact, if the underlying physics governing the \spell{behaviour}{behavior}
of the system  is ``quantum'' (in the sense that it can be based on entangled
quantum  states, rather than correlated random variables) then
\spell{behaviour}{behavior} can occur  that is impossible in the classical 
case.
This is a natural way of interpreting Bell's theorem \cite{Bell,CHSH}.
To formalize---and later generalize---this, we shall define 
{\em quantum measurement scenarios\/} and {\em (classical) local hidden
variable schemes}.

\PRL{}{\section{Definitions and preliminary results}}

Define a {\em quantum measurement scenario\/} as a triple of 
the form $(\entang,M_A,M_B)$, where $\entang$ is a 
bipartite quantum state, $M_A$ is a set of measurements on the 
first component, and $M_B$ is a set of measurements on the second 
component.

It is convenient to parametrize the simplest von Neumann measurements 
on individual qubits by points on the unit circle (more general von 
Neumann measurements, which involve complex numbers, are considered 
later in this paper).
Let the parameter $x \in [0,2\pi)$ denote a measurement with respect 
to the operator 
\begin{equation}
\label{meas1}
R(x) = 
\pmatrix{\cos x & \ \ \ \sin x \cr 
         \sin x & -\cos x \cr}
\end{equation}
(whose eigenvectors are 
$\cos({x \over 2})\ket{0} + \sin({x \over 2})\ket{1}$ 
and 
$\sin({x \over 2})\ket{0} - \cos({x \over 2})\ket{1}$).

Consider the case of a pair of qubits in the Bell state 
$\bell = \shalf\ket{0}\ket{0} + \shalf\ket{1}\ket{1}$.
[Our results are written for such states, but can be modified to 
apply to any of the other Bell states, including the 
Einstein-Podolsky-Rosen singlet state 
$\epr = \shalf\ket{0}\ket{1} - \shalf\ket{1}\ket{0}$.]
Let $x, y \in [0,2\pi)$ be the respective measurement parameters 
of the two components and let $a, b \in \01$ be the respective 
outcomes.
Then the joint probability distribution of these outcomes is given as:

\begin{center}
\begin{tabular}{c|c|c|}
\multicolumn{1}{c}{} &
\multicolumn{1}{c}{$\Pr[b=0]$} &
\multicolumn{1}{c}{$\Pr[b=1]$} \vspace*{1mm}\\
\cline{2-3}
& & \vspace*{-3mm} \\
$\Pr[a=0]$ & $\half \cos^2({x-y \over 2})$ & $\half \sin^2({x-y \over 2})$ \\
[1ex]
\cline{2-3}
& & \vspace*{-3mm} \\
$\Pr[a=1]$ & $\half \sin^2({x-y \over 2})$ & $\half \cos^2({x-y \over 2})$ \\
[1ex]
\cline{2-3}
\end{tabular}
\end{center}

\smallskip

Two simple but noteworthy examples of bipartite quantum measurement 
scenarios with the Bell state $\bell$ are:
\begin{description}
\item[Example 1:]
$(\bell,M_A,M_B)$, where $M_A = M_B = \{0,\pihalf\}$.
\item[Example 2:]
$(\bell,M_A,M_B)$, where
$M_A = \{-{\pi \over 8}, {3 \pi \over 8}\}$ and
$M_B = -M_A = \{{\pi \over 8}, -{3 \pi \over 8}\}$.
\end{description}
In both examples, each individual outcome is a uniformly distributed 
bit regardless of the measurements.
In Example 1, if the two measurements are the same then the outcomes are 
completely correlated; whereas, if the two measurements are different, 
the outcomes are completely independent.
In Example 2, the two outcomes are equal with probability 
$\sin^2({\pi \over 8})$ if $x = -y = +{3 \pi \over 8}$; and 
with probability $\cos^2({\pi \over 8})$ otherwise.
These examples are interesting in the context of local hidden variable 
schemes, which are defined next.

Intuitively, we are interested in classical devices that simulate 
bipartite quantum measurement scenarios to varying degrees, and such 
devices are naturally explained as local hidden variable schemes.
To define a {\em local hidden variable scheme}, it is convenient 
to view it as a two-party procedure whose execution occurs in two stages: 
a {\em preparation stage\/} and a {\em measurement stage}.
For ease of reference, call the two parties Alice and Bob.
During the preparation stage, {\em local hidden variables\/} $u$ for Alice 
and $v$ for Bob are determined by a classical random process.
During this stage, arbitrary communication can occur between the two 
parties, so $u$ and $v$ may be arbitrarily correlated.
During the measurement stage, measurements $x$ and $y$ are given to Alice 
and Bob (respectively), who produce outcomes $a = A(x,u)$ and $b = B(y,v)$ 
(respectively).
During this stage, no communication is permitted between the parties, 
which is reflected by the fact that the value of $A(x,u)$ is 
independent of the value of $y$ (and vice versa).

A local hidden variable scheme {\em simulates\/} a 
measurement scenario $(\entang,M_A,M_B)$ if, for any 
$x \in M_A$ and $y \in M_B$, the outputs produced by Alice and Bob, 
(namely, $a$ and $b$ respectively), have exactly the same bivariate 
distribution as the outcomes of the quantum measurement scenario as 
dictated by the laws of quantum physics.

The measurement scenario in Example~1 is easily 
simulatable by the following local hidden variable scheme.
Let $u$ and $v$ each consist of a copy of the {\em same\/} uniformly 
distributed two-bit string.
Then let Alice and Bob each output the first bit of this string if their 
measurement is 0 and the second bit if their measurement is $\pihalf$.
On the other hand, for the measurement scenario of Example~2, it turns 
out that {\em there does not exist\/} a local hidden variable scheme 
that simulates it \cite{CHSH}.

Now, we consider a more powerful classical instrument for simulating 
measurement scenarios.
Define a local hidden variable scheme {\em augmented by $k$ bits of 
communication}, as follows.
Informally, it is a local hidden variable scheme, except that the 
prohibition of communication between the parties during the measurement 
stage is relaxed to a condition that allows up to $k$ bits of communication 
(but no more).
More formally, a local hidden variable scheme augmented by $k$ bits of 
communication, has a preparation stage where random variables $u$ and $v$ 
for Alice and Bob are determined and during which arbitrary communication 
is permitted between the two parties.
Then there is a measurement stage which begins by measurements $x$ and $y$ 
being given to Alice and Bob (respectively).
Then one party computes a bit (as a function of his/her measurement 
and local hidden variables) which is sent to the other party.
This constitutes one {\em round\/} of communication.
Then again one party (the same one or a different one) computes a bit 
(as a function of his/her measurement, local hidden variables, and any 
data communicated from the other party at previous rounds) and sends it 
to the other party.
And this continues for $k$ rounds, after which Alice and Bob output 
bits $a$ and $b$ (respectively).

For example, for the measurement scenario of Example~2, a local hidden 
variable scheme augmented with one single bit of communication can 
simulate it.
This is a consequence of the following more general result, whose
easy proof \PRL{is left as an exercise.}{we include for completeness.}

\smallskip \noindent
{\bf Theorem~1.}
{\sl For any quantum measurement scenario $(\entang,M_A,M_B)$, 
there exists a local hidden variable scheme augmented with $\log_2(|M_A|)$ 
bits of communication (from Alice to Bob) that exactly simulates it.}\PRL{}{

\smallskip \noindent
{\bf Proof.}
First note that, if we allow $\log_2(|M_A|)$ bits of communication from 
Alice to Bob {\em and\/} $\log_2(|M_B|)$ bits of communication from Bob to 
Alice then it is trivial to simulate the quantum measurement scenario.
With this much communication, Alice can obtain $y$ and Bob can obtain $x$, 
which effectively defeats any ``nonlocality'' in the scenario.
More precisely, during the preparation stage, Alice and Bob can construct 
$|M_A| \cdot |M_B|$ random variable pairs, $(a^{(x,y)},b^{(x,y)})$, one 
for each value of $x \in M_A$ and $y \in M_B$.
Each such random variable pair would specify the values of the outcomes 
of Alice and Bob for the given values of $x$ and $y$, with the appropriate 
correlation.
During the measurement stage, after the communication of $x$ and $y$ 
between them, Alice and Bob can simply output $a^{(x,y)}$ and $b^{(x,y)}$ 
(respectively).

To obtain a protocol in which only $\log_2(|M_A|)$ bits of communication from
Alice to Bob occurs, note that the unconditional probability 
distribution of $a^{(x,y)}$ (the output of Alice when the measurements 
are $x$ and $y$) is independent of the value of $y$.
This is because the distribution of $a^{(x,y)}$ is completely determined 
by $x$ and the reduced density matrix of $\entang$ with the second 
component traced out ($\Tr_B(\entang)$), and this quantity is 
independent of~$y$.
Therefore, the local hidden variables can be set up as follows.
For each $x \in M_A$, $a^{(x)}$ is sampled according to the appropriate 
probability distribution, and then, for each $x \in M_A$ and $y \in M_B$, 
$b^{(x,y)}$ is sampled according the appropriate conditional probability 
distribution (conditioned on the value of $a^{(x)}$) in order to produce 
the correct bivariate distribution for $(a^{(x)},b^{(x,y)})$.
Then, during the measurement stage it suffices for Alice to send $x$ to Bob, 
and for Alice and Bob to output $a^{(x)}$ and $b^{(x,y)}$ (respectively).
\hfill \qed}
\smallskip

We shall see that in some cases the upper bound of Theorem~1 is 
asymptotically tight while in other cases it is not.
In \PRL{what follows}{the sections that follow}, we focus on the case 
of a single Bell state and the case of $n$ Bell states, and provide a new 
upper or lower bound in each case.

\PRL{}{\section{The case of a single Bell state}}

Consider the case of a single Bell state 
$\bell = \shalf\ket{0}\ket{0} + \shalf\ket{1}\ket{1}$,
but where the sizes of $M_A$ and $M_B$  may be arbitrarily large.
By Theorem~1, we only obtain an upper bound of $\log_2(|M_A|)$ bits for the 
amount of communication necessary for an augmented local hidden variable 
scheme to simulate it.
In the case where $M_A$ and $M_B$ are each the entire interval $[0,2\pi)$, 
this communication upper bound would be infinite.
If only a finite number, $k$, bits of communication are permitted then one 
alternative that might seem reasonable is for Alice to send 
$x^{\prime}$, a $k$-bit approximation of $x$, to Bob.
The protocol for Alice and Bob would be along the lines of the one 
\PRL{implicit}{} in Theorem~1, but using $x^{\prime}$ in place of $x$.
This would clearly not produce an exact simulation for a general 
$x \in [0,2\pi)$, but it would produce an {\em approximation\/} that improves 
as $k$ increases.
Is this the best that can be done with $k$ bits of communication?
The next theorem demonstrates that it is possible to obtain
an {\em exact\/} simulation for any $x, y \in [0,2\pi)$ with only a
{\em constant\/} number of bits of communication.

\smallskip \noindent
{\bf Theorem~2.}
{\sl For the quantum measurement scenario 
$(\bell,M_A,M_B)$ with 
$\bell = \shalf\ket{0}\ket{0} + \shalf\ket{1}\ket{1}$ 
and $M_A = M_B = [0,2\pi)$, there exists a local hidden variable 
scheme augmented with four of bits of communication (from Alice to Bob) 
that exactly simulates it.}

\smallskip \noindent
{\bf Proof.}
The local hidden variables are $c \in \01$ and 
$\theta \in [0,{3 \pi \over 5})$, and both are uniformly distributed.

For $j \in \{0,1,\ldots,9\}$, define $\alpha_j = {j \pi \over 5}$.
It is useful to view $\alpha_0,\alpha_1,\dots,\alpha_{9}$ as ten 
equally-spaced points on the unit circle.
Define the $j^{\mbox{\scriptsize th}}$ {\em $\alpha$-slot\/} as the 
interval $[\alpha_j,\alpha_{(j+1) \bmod 10})$.
Also, define 
$\beta_0 = \alpha_0 + \theta$, 
$\beta_1 = \alpha_3 + \theta$, and 
$\beta_2 = \alpha_6 + \theta$ and 
$\gamma_0 = \alpha_5 + \theta$, 
$\gamma_1 = \alpha_8 + \theta$, and 
$\gamma_2 = \alpha_1 + \theta$ 
(where the addition is understood to be modulo $2 \pi$).
Define the $j^{\mbox{\scriptsize th}}$ {\em $\beta$-slot\/} as the 
interval $[\beta_j,\beta_{(j+1) \bmod 3})$, and the 
$j^{\mbox{\scriptsize th}}$ {\em $\gamma$-slot\/} as the 
interval $[\gamma_j,\gamma_{(j+1) \bmod 3})$.

The protocol starts by Alice sending Bob information specifying the 
$\alpha$-slot, $\beta$-slot, and $\gamma$-slot in which $x$ is located.
Note that these slots partition the unit circle into sixteen intervals, 
so Alice can convey this information by sending four bits to Bob.
Then Alice outputs the bit $c$.

The full procedure for Bob is summarized below, but, in order to explain 
the idea behind it, it is helpful to first consider the special case 
where $y$ is in the 
2$^{\mbox{\scriptsize nd}}$ $\alpha$-slot and the $\alpha$-slot number 
of $x$ is within two of that of $y$ (in other words, the $\alpha$-slot 
number of $x$ is in $\{0,1,2,3,4\}$).
Note that these conditions depend on the values of $x$ and $y$ only 
(and not on the values of the local hidden variables).
Also, these conditions imply that $|x-y| \le {3 \pi \over 5}$.
In this case, Bob does the following.
If the $\beta$-slots of $x$ and $y$ are the same then Bob outputs $c$.
If the $\beta$-slots of $x$ and $y$ are different then exactly one 
$\beta_k$ is between $x$ and $y$.
Let $u = |y - \beta_k|$.
Then Bob's procedure is to output $c$ with probability 
$1 - {3 \pi \over 10}\sin(u)$.

To \spell{analyse}{analyze} the stochastic \spell{behaviour}{behavior} of
this procedure (still in the special case), let $r = |x-y|$ and note that the
probability of $x$ and $y$ being  in different $\beta$-slots is ${5r \over
3\pi}$. Also, conditional on $x$ and $y$ being in different $\beta$-slots, the 
probability distribution of the position of the $\beta_k$ between $x$ and 
$y$ is uniform.
Therefore, 
\begin{eqnarray}
\Pr[a=b] & = & 
{\textstyle{\left(1-{5 r \over 3 \pi}\right) 
+ \left({5 r \over 3 \pi}\right)\left({1 \over r}\right)}}
\int_0^r {\textstyle{\left(1-{3 \pi \over 10}\sin(u)\right)}}du \nonumber \\
& = & \half (1 + \cos(r)) \nonumber \\
& = & \cos^2({\textstyle{r \over 2}}),
\end{eqnarray}
which is exactly what is required.

The procedure for Bob in the above special case can be generalized to 
apply to the other possible cases by considering various similarities 
and symmetries among the cases.
\PRL{Once Bob obtains information from Alice that specifies the
$\alpha$-slot, $\beta$-slot, and $\gamma$-slot of $x$,
he simply has to apply the following procedure:}%
{First note that the above procedure actually works in all cases where 
the $\alpha$-slot number of $y$ is in $\{2,3,4,5,6\}$ and the 
$\alpha$-slot number of $x$ is within two of that of $y$.
This is because, in these cases, the interval between $x$ and $y$ 
(of length $\le {3 \pi \over 5}$) lies entirely within the interval 
$[0,{9 \pi \over 5})$ and $\beta_0, \beta_1, \beta_2$ are uniformly 
distributed points spaced ${3 \pi \over 5}$ apart in this interval.

Now, consider the cases where the $\alpha$-slot number of $y$ is 
in $\{7,8,9,0,1\}$ and the $\alpha$-slot number of $x$ is still 
within two of that of $y$.
In these cases, the interval containing $x$ and $y$ may not lie 
entirely within $[0,{9 \pi \over 5})$, and so the distribution of 
$\beta_0, \beta_1, \beta_2$ may no longer satisfy the relevant 
properties.
To avoid this problem, Bob applies the above procedure with 
$\gamma_0, \gamma_1, \gamma_2$ substituted in place of 
$\beta_0, \beta_1, \beta_2$.
This works because $\gamma_0, \gamma_1, \gamma_2$ are uniformly 
distributed points spaced ${3 \pi \over 5}$ apart in the interval 
$[\pi,{4 \over 5}\pi)$ (taken {\em clockwise\/}) and the interval 
containing $x$ and $y$ is within this interval.

The above covers all cases where the $\alpha$-slot number of $x$ is 
within two of that of $y$.
To handle the remaining cases, Bob works with $y^{\prime} = y + \pi$ 
(whose $\alpha$-slot number will then be within two of that of $x$) 
instead of $y$.
Let $r^{\prime} = |x-y^{\prime}|$.
Then, since $\cos^2({r^{\prime} \over 2}) = \sin^2({r \over 2})$, 
Bob will obtain the required distribution if he applies the above 
procedure but negates his output bit.

In summary, Bob's procedure after obtaining information specifying 
the $\alpha$-slot, $\beta$-slot, and $\gamma$-slot of $x$ from 
Alice is:}

\smallskip

\ni {\bf if the difference between the $\alpha$-slot numbers of $x$ 
             and $y$ is more than two then} \nl
\ni \ii {\bf set $y$ to $y + \pi$} \nl
\ni \ii {\bf set $c$ to $\neg c$} \nl
\ni {\bf if the $\alpha$-slot number of $y$ is in $\{7,8,9,0,1\}$ then} \nl
\ni \ii {\bf set $\beta_0, \beta_1, \beta_2$ to 
                     $\gamma_0, \gamma_1, \gamma_2$} \nl
\ni {\bf if $x$ and $y$ are in the same $\beta$-slot then} \nl
\ni \ii {\bf output $c$} \nl
\ni {\bf else there exists a $\beta_k$ between $x$ and $y$} \nl
\ni \ii {\bf set $u$ to $|y-\beta_k|$} \nl
\ni \ii {\bf output $c$ with probability 
            $1 - {3 \pi \over 10}\sin(u)$}

\hfill \qed
\smallskip

Theorem~2 applies to all measurements with respect to operators of the 
form given in Eq.~(\ref{meas1}).
The most general possible von Neumann measurement on an individual qubit 
can be parametrized by $(x,\xx) \in [0,2\pi) \times [0,2\pi)$ and taken 
with respect to the operator
\begin{equation}
S(x,\xx) = 
\pmatrix{\ \ \ \ \ \cos x & e^{-i \xx} \sin x \cr 
         e^{i \xx} \sin x & \ \ \ \   -\cos x \cr}
\end{equation}
(whose eigenvectors are 
$\cos({x \over 2})\ket{0} + e^{i\xx}\sin({x \over 2})\ket{1}$ 
and 
$\sin({x \over 2})\ket{0} - e^{i\xx}\cos({x \over 2})\ket{1}$).
If Alice and Bob make such measurements with respective parameters 
$(x,\xx)$ and $(y,\yy)$ and $a$ and $b$ are the respective outcomes 
then $\Pr[a=0] = \Pr[b=0] = \half$ and 
\begin{eqnarray}
\label{corr2}
\Pr[a=b] & = & 
\PRL{
{\textstyle{\cos^2({\xx + \yy \over 2})\cos^2({x-y \over 2})}} \nonumber \\
& & {\textstyle{+ \sin^2({\xx + \yy \over 2})\cos^2({x+y \over 2}).}}
}{
{\textstyle{\cos^2({\xx + \yy \over 2})\cos^2({x-y \over 2})
+ \sin^2({\xx + \yy \over 2})\cos^2({x+y \over 2}).}}
}
\end{eqnarray}

\smallskip \noindent
{\bf Theorem~3.}
{\sl For the quantum measurement scenario 
$(\bell,M_A,M_B)$ with 
$\bell = \shalf\ket{0}\ket{0} + \shalf\ket{1}\ket{1}$ 
and $M_A = M_B = [0,2\pi) \times [0,2\pi)$, there exists a local hidden 
variable scheme augmented with eight bits of communication (from Alice 
to Bob) that exactly simulates it.}
\PRL{\looseness=-1}{}

\smallskip \noindent
{\bf Proof.}
The local hidden variable scheme consists of two executions of the 
four-bit protocol of Theorem~2.
In the first execution, Alice and Bob use measurement parameters $\xx$ 
and $-\yy$ to obtain output bits $\aa$ and $\bb$
\PRL{such that}{(respectively) such that}
\PRL{\looseness=-1}{}
\begin{equation}
\label{corr2.1}
{\textstyle{\Pr[\aa=\bb] = \cos^2({\xx+\yy \over 2}).}}
\end{equation}
In the second execution, Alice and Bob use measurement parameters 
$(-1)^{\aa} x$ and $(-1)^{\bb} y$ to obtain their final output bits 
$a$ and $b$ (respectively).
Note that 
\begin{equation}
\Pr[a=b] = \cases{\cos^2({x-y \over 2}) & if $\aa = \bb$ 
                  \vspace*{1mm}\cr
                  \cos^2({x+y \over 2}) & if $\aa \neq \bb$,\cr}
\end{equation}
which, combined with Eq.~(\ref{corr2.1}), implies Eq.~(\ref{corr2})%
\PRL{.}{ as required.}
\hfill \qed

\smallskip

\PRL{}{We do not know whether a similar result holds in the case 
of quantum measurements that are more general than von Neumann 
measurements (e.g.\ positive operator valued measures).}

\PRL{}{\section{The case of {\boldmath $n$} Bell states}}

\PRL{Consider the tensor product of $\bell$ with itself $n$ times,
i.e.\ the case of $n$ Bell states.}
{Consider the case of $n$ Bell states, i.e.\ the tensor product of 
$\bell$ with itself $n$ times.}
This state can be written as $\belln = {1 \over \sqrt{2^n}} 
\sum_{i \in \01^n} \ket{i}\ket{i}$.
Theorem~3 implies that any $n$ {\em independent\/} von Neumann 
measurements performed on the $n$ Bell states can be simulated by a
local hidden variable scheme augmented with $8n$ bits of communication.
In the case of {\em coherent\/} measurements on such a state, 
the exact simulation cost can be much larger, as
shown by the following theorem.

\smallskip \noindent
{\bf Theorem~4.}
{\sl There exists a pair of sets of measurements, $M_A$ and $M_B$ 
(each of size $2^{2^n}$) on $n$ qubits, such that, for the 
quantum measurement scenario $(\belln,M_A,M_B)$ with 
$\belln = {1 \over \sqrt{2^n}} 
\sum_{i \in \01^n} \ket{i}\ket{i}$, any local hidden variable scheme 
must be augmented with a constant times $2^n$ bits of communication 
in order to exactly simulate it.}

\smallskip \ni
{\bf Proof.}
The proof is based on connections between a measurement scenario and 
a communication complexity problem examined in \cite{BCW}.
We begin by defining a set of $2^{2^n}$ measurements, which we call 
{\em Deutsch-Jozsa\/} measurements, due to their connection with the 
algorithm in \cite{DJ}.
The measurements are parametrized by the set $\01^{2^n}$.
For a parameter value $z \in \01^{2^n}$, we index the bits of $z$ by 
the set $\01^n$.
That is, for $i \in \01^n$, $z_i$ denotes the 
``$\,i^{\mbox{\scriptsize th}\,}$'' bit of $z$.
The measurement on $n$ qubits corresponding $z \in \01^{2^n}$ is easily 
described as two unitary transformations followed by a measurement in 
the computational basis.
The first unitary transformation is a phase shift that maps 
$\ket{i}$ to $(-1)^{z_i}\ket{i}$ for each $i \in \01^n$.
The second unitary transformation is the $n$-qubit Hadamard 
transformation, which maps $\ket{i}$ to 
\begin{equation}
{\textstyle{1 \over \sqrt{2^n}}}\sum_{j \in \01^n}(-1)^{i \cdot j}\ket{j},
\end{equation}
where $i \cdot j$ is the inner product of the two $n$-bit strings 
$i$ and $j$ 
(that is, $i \cdot j = i_0 j_0 + i_1 j_1 + \cdots + i_{n-1} j_{n-1}$).
These two unitary transformations are followed by a measurement in the 
computational basis $\{\ket{i} : \mbox{$i \in \01^n$}\}$, 
yielding an outcome in $\01^n$.

Set $M_A = M_B = \01^{2^n}$, the set of Deutsch-Jozsa measurements.
We will now show that, for $x \in M_A$ and $y \in M_B$, the joint 
probability distribution of the outcomes $a$ and $b$ satisfies the 
following properties:
\begin{enumerate}
\item
If $x = y$ then $\Pr[a=b] = 1$.
\item
If the Hamming distance between $x$ and $y$ is $2^{n-1}$ then 
$\Pr[a=b] = 0$.
\end{enumerate}
To show this, consider the quantum state after the phase flips and 
Hadamard transformations have been performed, but before the measurement.
First, applying the phase flips to $\belln$ yields the state 
\begin{equation}
{\textstyle{1 \over \sqrt{2^n}}}
\sum_{i \in \01^n}(-1)^{x_i + y_i}\ket{i}\ket{i}.
\end{equation}
Next, after applying the Hadamard transformations, the state becomes 
\begin{equation}
\label{djstate}
{\textstyle{1 \over \sqrt{2^{3n}}}}
\sum_{j, k, i \in \01^n}
(-1)^{x_i + y_i + i \cdot (j \xor k)}\ket{j}\ket{k}
\end{equation}
(where $j \xor k$ is the bit-wise exclusive-or of $j$ and $k$).
To prove property~1, note that if $x = y$ then state (\ref{djstate}) 
becomes 
\[
{\textstyle{1 \over \sqrt{2^{3n}}}}
\sum_{j, k, i \in \01^n}
(-1)^{i \cdot (j \xor k)}\ket{j}\ket{k} 
= {\textstyle{1 \over \sqrt{2^n}}}
\sum_{i \in \01^n}\ket{i}\ket{i},
\]
so $\Pr[a=b] = 1$ when the measurement is performed.
To prove property~2, note that if the Hamming distance between $x$ 
and $y$ is $2^{n-1}$ then $x_i + y_i$ is even for $2^{n-1}$ values of 
$i$ and odd for $2^{n-1}$ values of $i$.
Therefore, the amplitude of any ket of the form $\ket{j}\ket{j}$ in 
state (\ref{djstate}) is
\begin{equation}
{\textstyle{1 \over \sqrt{2^{3n}}}}
\sum_{i \in \01^n}(-1)^{x_i + y_i} = 0,
\end{equation}
so $\Pr[a=b] = 0$.

Now we reduce a communication complexity problem in \cite{BCW} to the 
problem of designing an augmented local hidden scheme that satisfies 
properties 1 and 2.
The communication complexity problem (called $\mbox{\it EQ\/}^{\prime}$ 
in \cite{BCW}) is a restricted version of the ``equality'' problem, 
and is defined as follows.
Alice and Bob get inputs $x, y \in \01^{2^n}$ (respectively), and 
one of them (say, Bob) must output 1 if $x=y$ and 0 if the Hamming 
distance between $x$ and $y$ is $2^{n-1}$ (the output of Bob can be 
arbitrary in all other cases).
In \cite{BCW}, it is proven that any classical protocol that exactly 
solves this restricted equality problem requires $c 2^n$ bits of 
communication for some constant $c > 0$ 
(the proof is based on a combinatorial result in \cite{FR}).
Suppose that there exists a local hidden variable scheme augmented 
with $f(n)$ bits of communication that simulates the measurement 
scenario $(\belln,M_A,M_B)$.
Then one can use this to construct a protocol for restricted equality 
with $f(n) + n$ bits of communication as follows.
Alice and Bob first execute the protocol for 
$(\belln,M_A,M_B)$ and then Alice sends her output $a$ to 
Bob, who outputs 1 if $a=b$ and 0 if $a \neq b$.
It follows that $f(n) + n \ge c2^n$, so 
$f(n) \ge c2^n - n \ge c^{\prime}2^n$, for some $c^{\prime} > 0$ and 
sufficiently large $n$.
The theorem extends to all $n \ge 1$, possibly using a smaller constant
$c^{\prime\prime}$, because it follows from \cite{CHSH} that Example~2
cannot be simulated without communication.
\hfill \qed

\end{document}